\documentclass[12pt]{iopart}

\usepackage{graphicx}
\usepackage{bm}
\usepackage{braket}

\newcommand{\refsubfig}[2]{\fref{#1}(#2)} 
\newcommand{\CNOT}{\textsc{cnot}}
\newcommand{\SWAP}{\textsc{swap}}

\begin{document}

\title{Transport Implementation of the Bernstein-Vazirani Algorithm with Ion Qubits}
\author {S D Fallek, C D Herold, B J McMahon, K M Maller and J M Amini}
\address{Georgia Tech Research Institute, Atlanta, GA 30332, USA}
\author {K R Brown}
\address{School of Physics, Georgia Institute of Technology, Atlanta, Georgia 30332, USA}
\address{School of Chemistry and Biochemistry, Georgia Institute of Technology, Atlanta, Georgia 30332, USA}
\address{School of Computational Science and Engineering, Georgia Institute of Technology, Atlanta, Georgia 30332, USA}
\ead{Spencer.Fallek@gtri.gatech.edu}
\vspace{10pt}
\begin{indented}
\item[]\today
\end{indented}



\begin{abstract}
Using trapped ion quantum bits in a scalable microfabricated surface trap, we perform the Bernstein-Vazirani algorithm. 
Our architecture relies upon ion transport and can readily be expanded to larger systems. 
The algorithm is demonstrated using two- and three-ion chains. 
For three ions, an improvement is achieved compared to a classical
system using the same number of oracle queries. For two ions and one query, we correctly determine an unknown bit string with probability 97.6(8)\%. 
For three ions, we succeed with probability 80.9(3)\%.  

\end{abstract}


\vspace{2pc}
\noindent{\it quantum computing, quantum control, trapped ions, surface-electrode trap\/}


\section{Introduction}
The Bernstein-Vazirani (BV) algorithm
helped to solidify the potential promise of quantum computers. 
The algorithm was the first example of a quasi-polynomial speed-up 
over a probabilistic classical computer \cite{Bernstein1997,Simon1994}.
In this work, we demonstrate this elementary quantum algorithm in a scalable system with trapped ion qubits. 
With three qubits and one oracle query we can determine a hidden two-bit string with 
higher fidelity than the 50\% success rate of a classical algorithm. 
The experiment takes advantage of the transport capabilities of a 
microfabricated ion trap.  
Our work may serve as a blueprint for one node in a larger quantum system based upon ion transport \cite{Kielpinski2002}. 

The goal of the BV algorithm is to determine a secret string $s \in \{0,1\}^{n}$.
We are given access to an oracle which adds $s \cdot x$  (mod 2) to an ancilla bit $a$, 
where $x$ is a user provided $n$-bit string.
One can solve the classical version of the problem by querying the oracle $n$ times with 
$x = 2^{i},i\in\{0,1,\ldots,n-1\}$.
By examining the value of $a$ after each run, the user can determine the $i$th bit of $s$. 
However, using the BV quantum algorithm, $s$ can be determined with one oracle query.
By bracketing the oracle query with Hadamard transformations 
and preparing $a=1$ to generate a phase kickback \cite{Cleve1998},
the action of the oracle on the ancilla 
results in the bits of $s$ being mapped onto the bits in $x$.
To determine $s$, one need only measure the state of the data qubits (the ancilla qubit is returned to the $\ket{1}$ state). 
The algorithm provides a polynomial speed-up by a factor of $n$ for a single string $s$.
It can achieve a quasi-polynomial speed-up for the recursed problem \cite{Aaronson2002}.  

\begin{figure}[htb]\centering
	\includegraphics[width=0.7\textwidth]{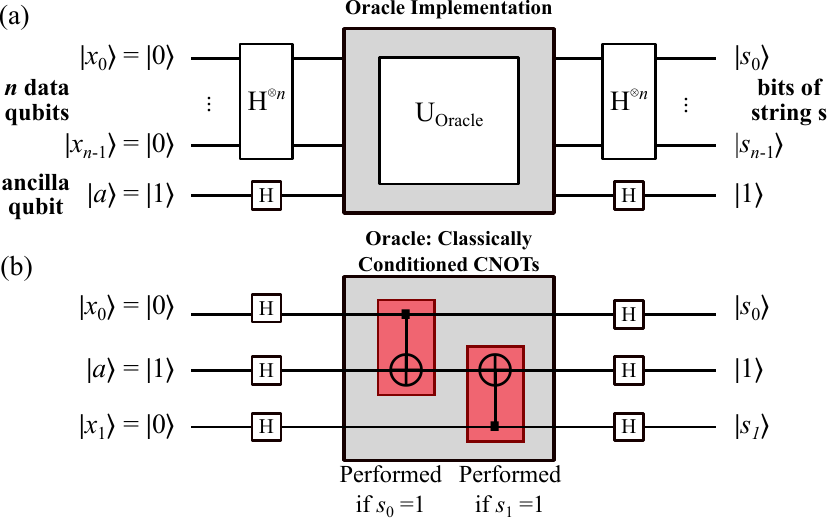}
	\caption{(a)~Circuit diagram for the BV algorithm. 
	The bits of $s$ are mapped to the data bits $x_{i}$ and can be determined
	in a single run of the algorithm.
	(b)~Our implementation of the BV algorithm with three ions. The ancilla
	qubit is in the center of the three-ion chain. The oracle is implemented
	via two \CNOT~gates. The execution of these gates is conditioned on 
	the classical bits in $s$. \label{fig:BVgen}}
\end{figure}

The circuit for the BV algorithm, shown in \refsubfig{fig:BVgen}{a}, 
contains the standard motif of a classical oracle transformed by Hadamard gates.
Up to the oracle, this is the same blueprint used for Deutsch and Jozsa's quantum algorithm 
and a key component of Grover's search algorithm \cite{Cleve1998}. 
If the input Hadamard gate on the ancilla qubit is not performed, the circuit
can be used to solve the learning parity with noise problem \cite{Cross2015,Riste2015}.

To implement the BV algorithm, we require an oracle that can perform the unitary 
$$U_{s} =   \sum_{x \in \{0,1\}^{n}} \sum_{a \in \{0,1\}} \ket{x}\bra{x} \otimes \ket{a \oplus (s \cdot x)} \bra{a},$$ 
where $\oplus$ is addition modulo 2.
We build this oracle from a series of \CNOT~gates acting on the ancilla qubit
and each data qubit $x_{i}$, as shown for $n=2$ in \refsubfig{fig:BVgen}{b}. 
The ancilla qubit is the target bit in each \CNOT.
To implement the dot product between $s$ and $x$, the application
of each \CNOT~ is conditioned on the classical bits in $s$, i.e.
we only perform a \CNOT~ between $x_{i}$ and $a$ if $s_{i} = 1$. 

\section{Experiment Overview}
We demonstrate the BV algorithm using a chain of $^{171}$Yb$^{+}$ ions in a microfabricated surface trap. The trap is described in \cite{Guise2015} 
and further details on $^{171}$Yb$^{+}$ trapping can be found in \cite{Olmschenk2007a}. 
In this system, single ions are sequentially loaded and then merged into a common harmonic well to build
a chain of the desired length. 

Following \cite{Hayes2010,Islam2014}, gates are performed via a Raman transition using a mode-locked, tripled YAG laser with a 355 nm wavelength and a repetition rate of $\nu_{r}=$ 119.12(1) MHz. 
Pairs of comb teeth separated by $106\times\nu_{r}$ span the 12.6 GHz qubit hyperfine splitting $\nu_{0}$, such that $\nu_{0}=(106\times\nu_{r}) +\Delta$. The necessary correction frequency, $\Delta\simeq16$~MHz, 
changes with drifts in $\nu_{r}$. To monitor this, we measure the beat between combteeth spaced by 14$\times\nu_r$. We feed this signal forward to an acousto-optic modulator (AOM) to add $(200+\Delta)$ MHz to one of our counter-propagating Raman beams. A second AOM shifts the frequency of the other Raman beam by 200 MHz for resonant carrier interactions
and by small offsets from 200 MHz to address motional modes.

While quantum algorithms have already been performed in three dimensional ion traps \cite{Gulde2003,Brickman2005,Schindler2013,Chiaverini2005,Debnath2016},
this is the first demonstration in a microfabricated planar trap.
The use of microfabrication permits scaling to larger algorithm demonstrations
by enabling repeatable production of many electrodes for ion transport, as well as 
integration of detection optics \cite{Merrill2011} and control electronics \cite{Guise2014}. 
These features will prove useful for implementing many-ion quantum systems \cite{Kielpinski2002,Monroe2013}.

Universal control over the ion chain is accomplished by addressing ions
pairwise, as described in \cite{Herold2016}. Briefly, we transport the ion chain
between a set of discrete gate locations. At each location, the gate lasers are
pulsed on, and one pair of ions is illuminated. Entanglement is provided
by nearest-neighbor M{\o}lmer-S{\o}rensen (MS) interactions
\cite{Sørensen2000}. To generate independent single qubit unitaries, we perform a
cascade: a unitary operation at each successive gate location across the chain.
Each unitary is composed of PB1-stabilized $\pi/2$ rotations $R_{\varphi}(\pi/2)$,
where $\varphi$ labels the axis of rotation in the equatorial plane of the Bloch sphere. 
The PB1 passband compensating sequence serves to alleviate amplitude errors on the 
two targeted ions
and suppress rotations on the others \cite{Wimperis1994,Brown2004,Brown2005}. We
have built a compiler that finds a minimum set of rotations to construct a
desired single qubit unitary. Additionally, the compiler compacts requested
single qubit gates into one unitary while accounting for previous operations
in the cascade. It also accounts for the off-resonant light shifts described below.

\section{System Improvements}
While we employ the same beam geometry as \cite{Herold2016}, system improvements
have led to the increased gate fidelities reported in \tref{tab:fidelitytab}.
One arm of our Raman beam pair now propagates through 
a 300 mm segment of photonic crystal fiber (LMA-PM-5, NKT Photonics)
before it is focused down to the trapping region. 
The mounted fiber provides improved beam mode quality and reduced 
spatial beam wander from free space propagation of the tripled YAG laser. 
The output of the fiber is focused to a waist ($1/e^2$ intensity half-width)
of roughly 5 $\mu$m, whereas the waist of the other arm of the Raman 
beam pair is 14 $\mu$m.
The $1/e$ half-width of the Raman interaction is largely set by the 5 $\mu$m beam,
which is slightly less than the ion spacing of the chain.
Without utilizing the hydrogen loading treatment outlined in \cite{Colombe2014}, 
we cannot propagate more than 20 mW of 355 nm light through the fiber
without solarization over a few days of operation. 
Thus, to perform pairwise single qubit $\pi/2$ rotations in 6 $\mu$s,
we send 1 mW through the fiber and 110 mW through the other free-space Raman beam. 
We propagate 2 mW through the fiber to implement MS gates in approximately $160~ \mu s$. 

\begin{table}[htb]
	\caption{\label{tab:fidelitytab} Percentage gate fidelities for two- and three-ion chains. 
	Single qubit (SQ) Clifford gate fidelities 
	are characterized via randomized benchmarking \cite{Knill2008}. MS fidelities are characterized according to the following procedure:
	all ions are prepared in the $\ket{0}$ state, a pairwise MS gate is performed, and then a parity measurement. 
	We report the fidelity of Bell state creation ($F_{\mathrm{Bell}}$) following the methods in \cite{Sackett2000}.
	We also report $P_{1}$, the unwanted $\ket{1}$ population in the untargeted ion, to indicate the level of crosstalk during MS gates. }
	\begin{indented}
	\lineup
	\item[]\begin{tabular}{ccccccc}
		\br
		Chain Length & Gate Type & Ion 0& & Ion 1& & Ion 2  \\ 
		\mr
		2 & SQ & 97.80(6)& & 98.47(5)& & --   \\
		3 & SQ & 97.4(1)& & 97.9(1)& & 98.6(1)  \\
		\br 
		&  & Target Pair& & $F_{\mathrm{Bell}}$ & & $P_{1}$  \\ 
		\mr
		 2 & MS & \it{01} & &96.1(8) & &--  \\
		 3 & MS &  \it{01}&& 89.6(9)&& 5.2(3)  \\
		 3 & MS & \it{12}&& 83.1(1.0)&& 10.2(6)  \\
		\br
	\end{tabular}
	\end{indented}
\end{table}

In \cite{Herold2016}, we describe errors due to imperfect overlap between the red and blue sideband beams during MS gates.
This is now mitigated by propagating both sideband beams
through the same fiber, improving the Bell-state fidelities between addressed ions as presented in \tref{tab:fidelitytab}.
The addition of the fiber also nearly halves the crosstalk during pairwise MS gates on three ions.  
This is exemplified by the population in $\ket{1}$ of the unaddressed ion after Bell state preparation ($P_{1}$ in \tref{tab:fidelitytab}).  

We also discuss errors from imperfect pairwise addressing during single qubit operations in \cite{Herold2016}. 
We employed the PB1 sequence in that work, however the suppression was not sufficient to neglect population transfer on ions adjacent to the addressed pair (``neighbor ions"). 
With the fiber in place, the tighter beam focus and improved mode quality reduce the Rabi rate 
of neighbor ions to less than 20\% of the addressed ions. 
Under these conditions, for each resonant PB1-stabilized $\pi/2$ rotation, we expect neighbor ions to experience a worst case infidelity of $1-F=7\times10^{-4}$ with respect to the identity operation.  
This suppression allows us to neglect these unwanted neighbor ion rotations in our compiled algorithm.  

Carrier transitions on a pair of ions still exhibit some crosstalk
with the rest of the chain. 
Each Raman beam can induce Raman transitions by itself due to interactions of pairs of comb teeth. 
The laser repetition rate was chosen such that these single-beam interactions are well detuned from the carrier transition.
However, off-resonant couplings introduce a non-negligible shift in the qubit energy splitting.
The magnitude of the dominant component of this light shift is
$\delta_{ac}=\Omega^{2}_{ac}/2 \Delta$, 
where $\Omega_{ac}$ is the Rabi frequency of the shifting laser field. 
In our experiment, $\Omega_{ac}$ depends on an ion's location in the beam, thus
we measure the shift for each ion at each gate location using Ramsey spectroscopy. 
For the free-space beam, we measure $\sim 650$ Hz shifts when an ion is targeted, $\sim 350$ Hz shifts when it is a neighbor, 
and $\sim 100$ Hz shifts when it is two locations away from the target pair. 
Shifts due to the fiber-coupled beam are small ($<60$ Hz) and localized onto the targeted ion pair. 

We model the total light shift as a constant detuning during gate pulses.
On target ions, the PB1 sequence corrects for the shifts well, and we expect PB1-stabilized $\pi/2$ rotation infidelities of $7\times 10^{-5}$.
For untargeted ions, the sequence cannot correct for the shift. However, the error is well approximated by a z-rotation 
$R_{z}(\phi)$, where $\phi = 2\pi \delta \tau_{\mathrm{PB1}}$,
$\delta$ is the measured frequency shift in Hz, and
the time per PB1 sequence, $\tau_{\mathrm{PB1}}$, is $102~ \mu$s. 
For ions well outside of the beam,
this approximation becomes exact. To incorporate these rotations in our gate compiler, 
anytime we perform a gate operation, 
we include $R_{z}(\phi)$ for each untargeted ion. 

\begin{figure}[h] \centering
	\includegraphics[width=0.7\textwidth]{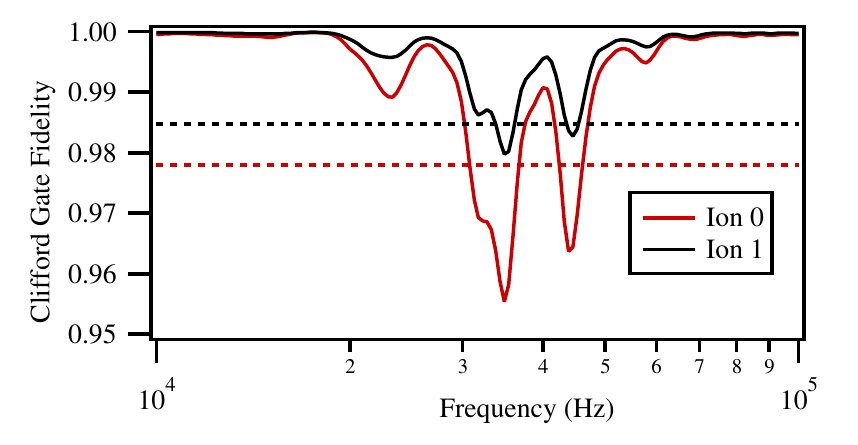}
	\caption{
	Simulated Clifford gate fidelity as a function of frequency of applied phase noise. 
	Each simulation is an average over 50 randomly generated randomized benchmarking sequences, 
	each with 15 gates per ion and a random starting phase of our noise source. 
	In our cascading architecture, Ion 0 is targeted for twice as many PB1 pulse sequences as Ion 1 on average.
	As expected, the simulation shows that the larger number of noisy gate operations leads to a higher error rate for Ion 0.  
	The dashed lines indicate the measured randomized benchmarking fidelities for each ion in a two-ion chain.  
	\label{fig:NoiseRB}}
\end{figure}

We believe that the primary error source for single qubit Clifford operations
is phase noise in the Raman interaction due to imperfect tracking of $\nu_{r}$. 
\Fref{fig:NoiseRB} shows results from numerical simulations of two-ion randomized 
benchmarking after fifteen Clifford operations per ion.
To match our physical implementation, the simulations incorporate each gate laser pulse, 
account for hardware programming delays, and include the time for ion transport. 
We plot single qubit Clifford gate fidelity as a function of the frequency of an applied phase noise. 
The phase noise is modeled at a single frequency and of modulation depth 0.02 radians.
Noise in the 20 kHz - 60 kHz frequency range is particularly detrimental.
The period of this noise ($16~\mu$s - $50~\mu$s) is on the order of $\tau_{\mathrm{PB1}}$.
We have observed noise near this sensitive band using the 
techniques described in \cite{Kotler2011}. 
However, technical limitations prevented us from measuring noise with frequency greater than $15$ kHz. 
In future work, we hope to characterize and mitigate these noise sources
for higher fidelity operation. 

\section{\label{sec:Alg}Algorithm Implementation}

\begin{figure}\centering
		\includegraphics[width=\textwidth]{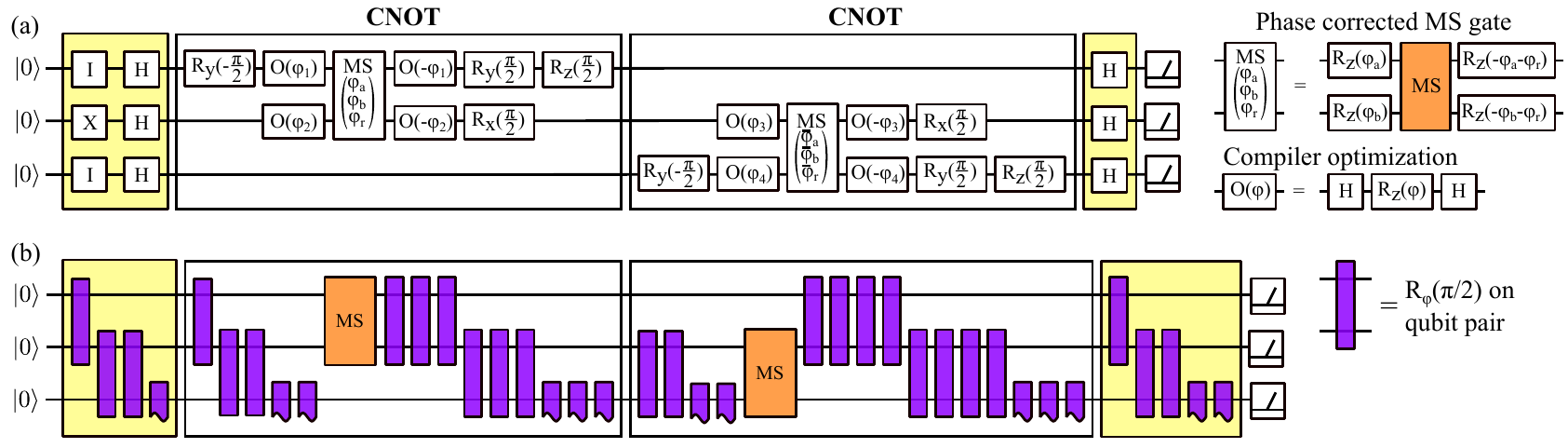}
		\caption{\label{fig:BVseq}
		(a)~Circuit diagram for a three-ion implementation of the BV algorithm. 
		The algorithm is broken up into four sections. The
		state preparation and analysis pulses are shown in the two yellow boxes.
		Conditional \CNOT~gates carry out the function of the oracle.
		Each \CNOT~contains a phase corrected MS gate, discussed in \sref{sec:Alg}.  		
 		Gate operations in each section are compiled to reduce the total number of operations. 
		However, gates across sections are not compiled. 
		(b)~Schematic of pairwise addressing for compiled BV. Each purple bar
		represents a PB1-stabilized $\pi/2$ pulse operation about an axis $\varphi$ in the equatorial plane.  
	}
\end{figure}

Each implementation of the algorithm begins with Doppler cooling and Raman sideband cooling.
After cooling, we measure an average temperature of $\overline{n}_\mathrm{COM}=0.6(1)$ 
quanta in the radial center of mass (COM) mode. 
All other radial modes are cooled to an average temperature below 0.1 quanta. 
The ions are then initialized to $\ket{0^{\otimes n+1}}$ via optical pumping. 
We follow this with a cascade of single qubit rotations to Hadamard transform each data qubit to $\ket{+}= \frac{1}{\sqrt{2}}(\ket{0}+ \ket{1})$
and the ancilla to $\ket{-}= \frac{1}{\sqrt{2}}(\ket{0} - \ket{1})$.
In the three-ion case, the ancilla qubit is the middle ion of the chain to allow for nearest-neighbor MS
gates, as shown in \fref{fig:BVseq}. To separate the input state preparation from the oracle implementation, these preparation gates 
are not compiled into any subsequent gates. 

To apply the oracle operation, \CNOT~gates are performed between each data ion and the ancilla,
conditioned on the value of $s$. \CNOT~ gates are built from MS gates with particular single qubit rotations before 
and after, as shown in \refsubfig{fig:BVseq}{a}. 
Due to increased axial confinement used during MS gates \cite{Herold2016}, we sample a different phase of our Raman beams than during single qubit gate operations. 
To negate this frame shift, we add phase corrections $R_{z}(\varphi_{a,b})$ and $R_{z}(-\varphi_{a,b}-\varphi_{r})$ to our compiled algorithm. 
The phase $\varphi_{r}$ accounts for light shifts accumulated during the MS gate. To accommodate these phase shifts in the fewest number of physical $\pi/2$ rotations, 
we insert optimization gates O($\varphi$), where $\varphi$ is a free parameter in the gate compilation. 
As described in \cite{Herold2016}, the addition of these
gates still produces the desired \CNOT. 

\CNOT~gates need to be turned on and off conditioned on the bits in $s$. 
If $s_{i} =0$, all the transport associated 
with each gate is performed, however the gate lasers are not turned on. 
This ensures algorithm honesty, as we do not optimize for any particular value of $s$.
Each \CNOT~ block is compiled separately from all other algorithm sections so that it can be turned on or off without effecting the rest of the circuit.

The oracle is followed by Hadamard gates on each ion and individual-ion state detection.
In total, the algorithm takes 3.9 ms for two ions and 9.8 ms for three ions.
The two-ion algorithm employs nine transport operations, 
and the three-ion algorithm requires nineteen.    
Each adiabatic transport is performed in $100~\mu s$.   

\section{Algorithm Output}
\begin{figure}[h] \centering
	\includegraphics[width=0.7\textwidth]{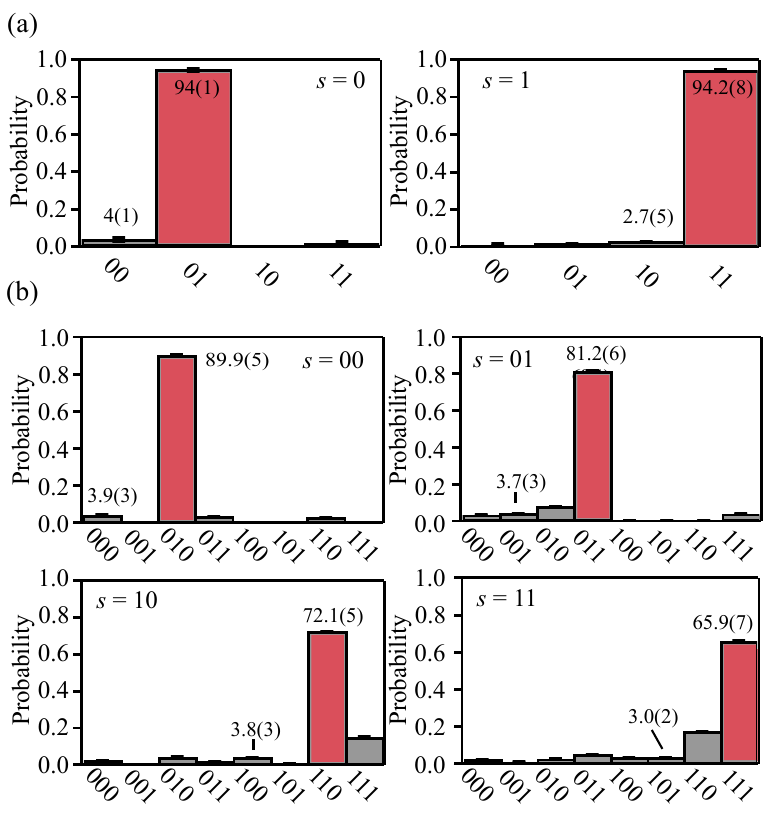}
	\caption{\label{fig:BV3}(a)~State populations after BV implementation on a two ion chain for different oracle inputs $s$. 
	The expected state $\ket{s}\otimes\ket{1}$ is marked in red.
	Values listed are the percentage occupation for the two states where $s$ was
	marked correctly, regardless of the state of the ancilla.
	(b)~Three ion implementation with $s$ containing two bits. 
	With the ancilla in the center of the chain the expected output state is now  $\ket{s_{0}}\otimes\ket{1}\otimes\ket{s_{1}}$.}
\end{figure}

To compare the BV algorithm to its classical counterpart, one can 
imagine that the user is given access to one classical query of the oracle 
which provides the value of a single bit in $s$.
Then he or she must guess the remaining bits.
Thus, for an $n$-bit secret string $s$, the user will guess the string correctly
with probability $2^{-n+1}$.
We implement the BV quantum algorithm for $n=1,2$ and provide state 
detection results in \fref{fig:BV3}. 

For $n=1$, a classical system can determine the single bit in $s$ correctly 100\% of the time. 
We find a quantum success probability of 97.6(8)\%, averaged over the two oracle states. This is taken by summing over
the states where the data bit $s$ is in the correct state, regardless of the 
state of the ancilla. 
In the $s=0$ case, we perform six PB1-stabilized $\pi/2$ rotations. Based upon randomized benchmarking sequences
of similar length, we expect the correct output state with fidelity of $95(3)\%$.
When $s=1$, we employ fifteen PB1-stabilized $\pi/2$ rotations and one MS gate.
Treating the error from the two gate types as uncorrelated, we expect 
an output state fidelity of $89(3)\%$. 
The quoted uncertainties in our predictions correspond to the spread of 
randomized benchmarking results. 
The results in both cases of \refsubfig{fig:BV3}{a} are consistent with their respective predictions
to within two standard deviations. 

For $n=2$, we correctly determine the bit string $s$ with probability 80.9(3)\%, averaged over the four oracle states. This is an improvement over the 
classical success rate of 50\% for a single query.
Furthermore, we can use information theory
to determine the information gained by one run of this algorithm.
We calculate the entropy reduction of one pass of the algorithm by examining the mutual information
between our algorithm output $x$ and a bit string $s$  \cite{Brickman2005,Nielsen2010}. 
Classically, as expected, the user
gains exactly one bit of information per oracle query.
Quantum mechanically we gain 1.10(3) bits of information.

In the three-qubit algorithm, our largest source of error is the MS gate on the $\it{12}$ pair.
This gate is the lowest fidelity operation in the system and causes significant crosstalk on
the untargeted ion. This manifests itself in the algorithm output as a large correlation 
between the $\ket{110}$ and $\ket{111}$ states when the gate is performed.
When $s=10$ there is $14.8(5)\%$ population in the $\ket{111}$ state, and
when $s=11$ there is $17.3(5)\%$ build up in $\ket{110}$. 
These populations have a drastic effect on algorithm fidelity and reduce
the target output states below 80\%. 
Preliminary work with the ``echo'' decoupling technique discussed in \cite{Herold2016} improves $F_{\mathrm{Bell}}$ to
$87.9(1.0)\%$ and reduces $P_{1}$ to 3.4(4)\%. The echo involves splitting the MS gate
in half, ensuring the spin-motion entanglement is zero after each half,
and inserting a $Y$ gate on the targeted (or untargeted) ions after each half.
At present, the additional errors introduced by the extra set of single qubit operations outweigh the improved decoupling.
We expect that with improved single qubit operation fidelities we can incorporate this technique 
in our algorithm. 

\section{Scaling}
One concern for scaling such an architecture is motional mode excitation 
due to anomalous heating and ion transport. 
We measure an anomalous heating rate of 50(6) quanta/s on the COM mode. 
Heating on other modes is less than 7 quanta/s.    
The MS two-qubit interaction and 
the PB1 pulse sequence were chosen to help 
mitigate gate errors due to increasing ion temperature. 
We do not believe that these errors 
are a current fidelity limitation. 

For MS gates, we couple mainly to the rocking mode to avoid the larger 
anomalous heating rate of the COM mode \cite{King1998}. 
We have simulated the density matrix evolution of a two-ion MS gate,
expanding the Hamiltonian out to third order in the Lamb-Dicke parameter. 
Following \cite{Sørensen2000}, we incorporated a coupling between vibrational states and a thermal reservoir. 
This predicts that heating on the COM mode will cause errors on the order of $10^{-4}$ per gate.
 
For single qubit operations, the PB1 sequence largely 
alleviates the carrier Rabi rate dependence on ion temperature. 
Simulations of a two-ion chain indicate that  we require $\overline{n}_\mathrm{COM}\geq5.5$ 
quanta to accumulate a sizable infidelity ($>10^{-4}$)
per PB1-stabilized $\pi/2$ rotation. After performing the two-ion BV algorithm with
transport, but with gate lasers off, we measure $\overline{n}_\mathrm{COM}=0.9(1)$.

In a three-ion chain, we require
$\overline{n}_\mathrm{COM}\geq 8.5$ quanta to exceed the same error threshold.  
The discrepancy compared to the two-ion case 
is due to the different Lamb-Dicke parameters. 
We measure $\overline{n}_\mathrm{COM}=1.6(1)$ quanta 
after the three-ion BV algorithm with transport. If we replace the algorithm
with an equivalent delay, $\overline{n}_\mathrm{COM}$ is consistent at 1.8(1) quanta. 
This affirms that the nineteen transport operations, each of about $5~ \mu$m,
are not a significant source of additional heating.

Thus, at our current heating rate, we do not expect that increasing ion temperature 
contributes largely to gate infidelities during the BV algorithm. 
However, as fidelities improve, scaling the system up may require implementation of proven technologies to reduce the heating rate \cite{Hite2012,McConnell2015}, 
to transport more quickly \cite{Walther2012}, and to allow for cooling after transport \cite{Lin2013}.

An additional impediment to scaling up is
that we are currently limited to nearest-neighbor \CNOT~gates. 
However, we can add \SWAP~ operations, built from three \CNOT s, to alleviate this restriction.
Another option is to employ the MS echo decoupling technique  
to generate entanglement between non-nearest neighbors \cite{Herold2016}. 
Finally, we could address the ions with two sets of Raman beams, each focused for individual ion addressing \cite{Debnath2016}. 
This would also allow two-qubit operations between any pair of ions
and would eliminate the need for a cascade during single qubit operations. 

\section{Outlook}
In this work, we have demonstrated a quantum algorithm in a microfabricated ion trap using transport-based ion addressing. Our three-ion implementation of the algorithm provides the user with more information than its classical
counterpart. We utilized temperature-insensitive gate operations to mitigate the effects of ion-heating. By demonstrating qubit control interlaced with up to nineteen transport operations,
we affirm the exceptional transport capabilities of surface electrode traps. 

\ack 
The authors thank J. True Merrill and Adam Meier for help with numerical simulations of 
randomized benchmarking. 
This material is based upon work supported by the Office of the Director of
National Intelligence (ODNI), Intelligence Advanced Research Projects Activity
(IARPA) under U.S. Army Research Office (ARO) contract W911NF1010231.  All
statements of fact, opinion, or conclusions contained herein are those of the
authors and should not be construed as representing the official views or
policies of IARPA, the ODNI, or the U.S. Government.

\section*{References} 

\bibliographystyle{iopart-num}
\bibliography{refsBViop}

\end{document}